\documentclass[aps,twocolumn,prd,showpacs,floatfix]{revtex4}
\usepackage{epsfig}
\usepackage{bm}
\usepackage{dcolumn}
\usepackage{multirow}
\usepackage{natbib}
\usepackage{amsmath,amsfonts,amssymb}
\usepackage{epsfig}
\usepackage{ifthen}
\usepackage{color}

\def\reff@jnl#1{{\rm#1\/}}
\def\aj{\reff@jnl{AJ}}                  % Astronomical Journal
\def\araa{\reff@jnl{ARA\&A}}            % Annual Review of Astron and Astrophys
\def\apj{\reff@jnl{ApJ}}                % Astrophysical Journal
\def\apjl{\reff@jnl{ApJ}}               % Astrophysical Journal, Letters
\def\apjs{\reff@jnl{ApJS}}              % Astrophysical Journal, Supplement
\def\ao{\reff@jnl{Appl.Optics}}         % Applied Optics
\def\apss{\reff@jnl{Ap\&SS}}            % Astrophysics and Space Science
\def\aap{\reff@jnl{A\&A}}               % Astronomy and Astrophysics
\def\aapr{\reff@jnl{A\&A~Rev.}}         % Astronomy and Astrophysics Reviews
\def\aaps{\reff@jnl{A\&AS}}             % Astronomy and Astrophysics, Supplement
\def\azh{\reff@jnl{AZh}}                % Astronomicheskii Zhurnal
\def\baas{\reff@jnl{BAAS}}              % Bulletin of the AAS
\def\jrasc{\reff@jnl{JRASC}}            % Journal of the RAS of Canada
\def\memras{\reff@jnl{MmRAS}}           % Memoirs of the RAS
\def\mnras{\reff@jnl{MNRAS}}            % Monthly Notices of the RAS
\def\pra{\reff@jnl{Phys.Rev.A}}         % Physical Review A: General Physics
\def\prb{\reff@jnl{Phys.Rev.B}}         % Physical Review B: Solid State
\def\prc{\reff@jnl{Phys.Rev.C}}         % Physical Review C
\def\prd{\reff@jnl{Phys.Rev.D}}         % Physical Review D
\def\prl{\reff@jnl{Phys.Rev.Lett}}      % Physical Review Letters
\def\pasp{\reff@jnl{PASP}}              % Publications of the ASP
\def\pasj{\reff@jnl{PASJ}}              % Publications of the ASJ
\def\qjras{\reff@jnl{QJRAS}}            % Quarterly Journal of the RAS
\def\skytel{\reff@jnl{S\&T}}            % Sky and Telescope
\def\solphys{\reff@jnl{Solar~Phys.}}    % Solar Physics
\def\sovast{\reff@jnl{Soviet~Ast.}}     % Soviet Astronomy
\def\ssr{\reff@jnl{Space~Sci.Rev.}}     % Space Science Reviews
\def\zap{\reff@jnl{ZAp}}                % Zeitschrift fuer Astrophysik
\def\nat{\reff@jnl{Nature}}             % Nature 

\def\pr{{\rm Pr}}
\def\data{\boldsymbol{d}}
\def\datai{\boldsymbol{d}_i}
\def\model{\mathsf{H}}
\def\modelA{\mathsf{H}_0}
\def\modelB{\mathsf{H}_1}
\def\Iratio{{\rm R}}
\def\rmd{{\rm d}}
\def\rmmax{{\rm max}}
\def\CMB{{\sc CMB}}
\def\SN{{\sc SN}}
\def\SDSS{{\sc SDSS}}

\begin{document}

\title[Dataset comparison]{Bayesian evidence as a tool
  for comparing datasets}

\author{Phil Marshall} 
\affiliation{Kavli Institute for Particle Astrophysics and Cosmology, Stanford University, USA}
\author{Nutan Rajguru}
\affiliation{Astrophysics Group, Cavendish Laboratory, Madingley Road,
   Cambridge,UK }
\author{An\v{z}e Slosar}
\affiliation{Faculty of Mathematics and Physics, University of Ljubljana, Slovenia}
\date{\today}

%-----------------------------------------------------------------------

\begin{abstract}
We introduce a new conservative test for quantifying the consistency of
two or more datasets. The test is based on the Bayesian answer to the question,
``How much more probable is it that all my data were generated from
the same model system than if each dataset were generated
from an independent set of model parameters?''.   We make explicit the
connection between evidence ratios and the differences in
peak chi-squared values, the latter of which are more widely used and
more cheaply calculated. Calculating evidence ratios for three
cosmological datasets (recent CMB data (WMAP, ACBAR, CBI, VSA), SDSS
and the most recent SNe Type 1A data) we find that concordance is
favoured and the tightening of constraints on cosmological parameters
is indeed justified. 
\end{abstract}


\topmargin-1cm

\maketitle

%-----------------------------------------------------------------------

\section{Introduction}

The apparent mutual agreement of a wide range of cosmological
observations has led to the current climate of ``concordance'' in
cosmology \citep{2003ApJS..148..175S,
2004MNRAS.353..747Rshort, 2004ApJ...609..498Rshort,
2002AAS...20114004K, 2004PhRvD..69j3501Tshort,
2004ApJ...607..665Rshort, 2002ApJ...581...20C,
2002MNRAS.335..432Vshort}.
The practice of combining independent datasets, by the multiplication of
their associated likelihood functions, in order to increase the
precision of the parameters of the world model is now standard, but
quantitative consistency checking is emphasised to a much lesser degree.
As all physicists will agree, accurate cosmology is preferable to
precision cosmology, and it is this that motivates this short
communication.

The purpose of this work is to demonstrate one application of Bayesian
model selection, that of checking that the far simpler model of a
universal set of parameters for modeling all datasets is justified by
the data themselves: in doing so we make the connection between the
Bayesian formulation of the problem and the pragmatic approach taken
at much lower computational cost by the experimental community.
In this work we show that, as is so often the
case, the standard approach is justified on the grounds of common
sense, and demonstrate the reduction of this common sense to
calculation via probability theory.

As usual, the route to model selection is via the Bayesian evidence. 
The evidence for a model $\model$ from data $\data$ is just the
probability $\pr(\data|\model)$, and can be calculated in principle by
marginalising the unnormalised posterior probability distribution
function over all $M$ parameters $\boldsymbol{\theta}$ in the model:
\begin{equation}
\pr(\data|\model) = \int
\pr(\data|\boldsymbol{\theta},\model)\pr(\boldsymbol{\theta}|\model)\,
d^M \boldsymbol{\theta}.
\end{equation}
In practice, calculating this integral is 
rarely feasible, but other techniques exist to provide estimates of the
evidence \citep[see, for example,][]{oruanaidh}. More detailed introductions to the evidence and its
central role in the problem of model selection are available elsewhere
\citep{sivia96,bishop} -- here we make the general remarks that the evidence increases
sharply with increasing goodness-of-fit, and decreases with increasing
model complexity (quantifying the principle of Occam's razor).  We show later explicitly how
these two aspects come to the surface and, for the specific case of
Gaussian measurement errors, result in model selection proceeding by 
the comparison of differences in the ubiquitous chi-squared statistic
with an ``Occam'' factor which takes the form of an effective number of
parameters. The more general approach advocated here is applicable to
any likelihood functions ($\pr(\data|\boldsymbol{\theta},\model)$), 
not just those having Gaussian form, and
takes into account the full extent of the pdf's involved. It is of
course also sensitive to the parameters' prior pdf 
($\pr(\boldsymbol{\theta}|\model)$): broader priors
represent more complex models and so naturally give lower evidence
values.  Evidence is the natural tool for comparing datasets in this
way: it enables us to quantify such questions as ``Is the mismatch
between two experiments large enough to warrant investigation into
possible sources of systematics or new physics?'' 

The simplest model for all the cosmological data in hand is that they
provide information on the same set of cosmological parameters: this is
the standard assumption made in all the joint analyses to date. Let
$\modelA$ represent the hypothesis that ``there is one set of
parameters that describes our cosmological model.'' In other words, we
believe that we understand both cosmology and our experiments to the
extent that there should be no further freedom beyond the parameters
specified. However, if we are interested in accuracy as well as
precision then we should take care to allow for systematic differences
between datasets: the most extreme case would be the one where the
observations were in such strong disagreement that they appeared to
give conflicting measurements of all the model parameters. In this case
one could consider the hypothesis $\modelB$ that ``there is a different
set of parameters for modeling each dataset.'' The conservatism of
such a model comparison exercise is readily apparent: the large
increase in model complexity incurred when moving from $\modelA$ to
$\modelB$ means that the joint analysis is intrinsically more
favourable. This means that any result in favour of $\modelB$ may be
taken as a clear indication of discord between the two experiments.
Note also that this test is easily done given that the evidence values
will have been calculated for alternative purposes, such as comparing
two physical models in the light of each dataset alone.

For checking dataset consistency then the quantity we should 
calculate is the ratio of
probabilities that each model is correct, given the data:
\begin{equation}
\frac{\pr(\modelA|\data)}{\pr(\modelB|\data)} =
\frac{\pr(\data|\modelA)}{\pr(\data|\modelB)}\cdot
\frac{\pr(\modelA)}{\pr(\modelB)}
\label{eq:AvsB}
\end{equation}
The calculable part of Equation~\ref{eq:AvsB} is the evidence ratio  
\begin{align} 
\Iratio &= \frac{\pr(\data|\modelA)}{\pr(\data|\modelB)} \notag \\
&= \frac{\pr(\data|\modelA)}{\prod_i \pr(\datai|\modelB)}
\label{eq:evratio}
\end{align} 
where in the second line we have assumed that the individual datasets
$\datai$ under analysis are independent. (The evidence integral
factors out since the independent likelihoods do, and also because each
likelihood depends only on its own subset of parameters.)
Interpretation of this evidence ratio is aided by
Equation~\ref{eq:AvsB}: for statement $\modelA$ to be more likely to
be true than statement $\modelB$, the product of $\Iratio$ and the
prior probability ratio must be greater than one. 
Suppose that an evidence ratio $\Iratio$ of 0.1 were found:  the
dataset combination ($\modelA$) can still be justified, but only if you
are willing to  take odds of ten to one on there being no
significant systematic errors in the system.  Blindly multiplying $N$
likelihoods together results, in general and approximately, in factors
of improvement in precision of $\sqrt{N}$: the evidence ratio gives an
indication of whether or not this improvement is justified, in the form
of an odds ratio (which enforces honesty through the
threat of bankruptcy).

Other criteria besides evidence have been used to compare different
models. Recently \citep{2004MNRAS.351L..49L} have proposed the Akaike
and Bayesian information criteria to carry out cosmological model
selection. These criteria are approximations to the full Bayesian
evidence under rather restrictive assumptions and thus fall under the
same framework. The posterior Bayes factors proposed by \cite{Murray}
and also discussed in \cite{2002PhRvD..66j3511L} can be used as an
alternative to evidence. This quantity is the Bayesian evidence with
the prior set to the posterior and can be readily estimated as an
average likelihood of the Markov Chain Monte Carlo chains. It has some
desirable properties, such as no prior dependence in the limit of
prior enclosing the entire volume of posterior. However, it has no
simple interpretation within Bayesian framework and will thus not be
discussed in this paper.  The use of the evidence itself as a model
selection tool has been growing in cosmology \citep[see
\emph{e.g.}][]{1996ApJ...471...24J,1998PhRvD..58h3004K,
  2002MNRAS.335..377H,2005astro.ph..4022T}.  Using evidence to
check dataset consistency has received much less attention.
Application to a particular problem of CMB map contamination can be
found in \cite{1998PhRvD..58h3004K}. In this work we construct a much
more general approach that can be applied to any setting in which a
given model is tested against more than one dataset. The price one has
to pay for this generality is that we are relatively insensitive to
any \emph{particular} inconsistency. We also aim to provide a
short tutorial, establishing the connection with
the more conventional $\chi^2$ statistics, followed by a 
simple analysis of current state-of-the-art experiments. 

%-----------------------------------------------------------------------

\section{Connection to $\chi^2$ analysis}
\label{sec:conn-chi2-analys}

Consider a general likelihood function
of some model parameter vector $\boldsymbol{x}$, which can be (for
reasons that will become apparent in a moment) rewritten as
\begin{equation}
L(\boldsymbol{x}) = L_\rmmax \hat{L}(\boldsymbol{x}),
\end{equation} 
where $L_\rmmax$ is the likelihood at the most likely point in the
parameter space and the dimensionless function $\hat{L}$ contains all
the likelihood shape information.
Assuming a uniform prior spanning between $-p$ and $p$ in each
direction,  where $p$ is large enough to encompass all region of high
likelihood,  gives the approximate evidence 
\begin{equation}
\tilde{E} \approx L_{\rmmax} \frac{\int \hat{L}
\rmd^{M}\boldsymbol{x}}{(2p)^M},
\end{equation} 
where $M$ is the number of parameters in the model.
If we identify the numerator of the above fraction with the
volume associated with the likelihood~$V_{L}$, and the denominator with
the available prior volume~$V_{\pi}$, we have  
\begin{equation}
\log{E} = \log{\left(\frac{V_{L}}{V_{\pi}}\right)} + \log{L_\rmmax}.
\label{evidbalance}
\end{equation} 
All the details of the
overlap between prior and likelihood is contained with in the volume
ratio, whereas the maximum likelihood value specifies the 
goodness-of-fit. Except when the posterior pdf's take simple analytic forms, this
volume factor must be calculated numerically and of course takes up much
of the effort in the evidence calculation.

In the case where the measurement errors are Gaussian, we can
write the evidence ratio used in this work in terms of 
the best-fit chi-squared
values that may be calculated during an analysis. 
It can be shown that

\begin{equation}
\log{R} = \log{\left(\frac{V_{12}V_{\pi}}{V_1 V_2}\right)} -
             \frac{1}{2} \Delta\hat{\chi}^2,
\end{equation} 
where $\Delta\hat{\chi}^2 = \hat{\chi}_{12}^2 - 
(\hat{\chi}_1^2 + \hat{\chi}_2^2)$. Defined this way, 
$\Delta\hat{\chi}^2$ is always positive (the goodness-of-fit cannot
decrease with the addition of the extra parameters) and we see that the
borderline case of $\log R = 0$ corresponds to the difference in
chi-squared between the two individual analyses and the joint fit being
equal to an effective number of parameters (difference in number of
degrees of freedom) given by the logarithm of the volume factor. 

Returning to the general case,
if we retain the assumption of a broad uniform prior, and 
if the likelihoods are well approximated by multivariate
Gaussians, then the volume factor can be calculated analytically:
in this case the $i^{\rm th}$
likelihood can be written as 
\begin{equation}
L_i \approx \hat{L}_i
\exp\left[-\frac{1}{2}(\boldsymbol{x}-\hat{\boldsymbol{x}}_i)^TF_i^{-1}(\boldsymbol{x}-\hat{\boldsymbol{x}}_i)\right],
\end{equation} 
where $F_i$ is the Fisher matrix. This gives, for the likelihood
volumes, 
 \begin{equation}
   V_i = \left(2\pi\right)^{M/2} \left| F_i \right|^{1/2}. 
 \end{equation}
In the joint analysis, combining two Gaussian likelihoods results in a
new Gaussian, centred at a correctly weighted mean of positions, but
whose shape is given simply by
\begin{equation}
  \hat{L_{12}} =
  \exp\left[-\frac{1}{2}(\boldsymbol{x}-\hat{\boldsymbol{x}}_i)^T\left(F_1
  + F_2\right)^{-1}(\boldsymbol{x}-\hat{\boldsymbol{x}}_i)\right],
\end{equation}
and therefore
\begin{equation}
  V_{12} = \left(2\pi\right)^{M/2} \left| F_1 + F_2 \right|^{1/2}. 
\end{equation}

Note that in this case, due to the high symmetry of the Gaussian 
approximation, the overlap integral $V_{12}$ is independent of
the distance between best fitting points. Therefore using $\Delta
\chi^2$ as a proxy for the Bayesian evidence change is valid when the
Gaussian approximation to the posterior is a good one. In the simple 
case where $F_1=F_2$ (a parallel degeneracy) and $V_{\pi} = (2p)^M$
again, the log evidence becomes:
\begin{equation}
%\log R = M \log\left(\frac{2p}{\sqrt{\pi|F|^{1/M}}}\right) - \frac{1}{2}\Delta \chi^2
\log R = \frac{1}{2} \left( M\log\left[\frac{4}{\pi}\frac{p^2}{|F|^{1/M}}\right]
- \Delta \chi^2 \right).
\end{equation}
The $\log$ term is typically of the order of unity: $|F|^{1/M}$ is the
geometrical average of the principal variances and hence $p^2
|F|^{-1/M}$ is the square of the 
ratio of the prior width to the characteristic likelihood width.  
Hence we recover the frequentist rule of
thumb that the increase in $\chi^2$ is justified if the number of
parameters drops by roughly the same number. However the evidence
considerations above allow this rule to be calibrated to take into
account both the prior information supplied and the (potentially
complex) shape of the likelihoods; in general, $V_{12}$ is not
independent of the individual peak positions, and so the simple
$\Delta \hat{\chi}^2$ procedure does not propagate all the information
contained within the likelihood functions.

%----------------------------------------------------------------------

\section{Comparing cosmological datasets}
\label{sect:data}
\subsection{ Datasets and method }

We use a version of the \texttt{CosmoMC} software package
\citep{2002PhRvD..66j3511L}, modified to calculate evidence by the
thermodynamic integration method. We obtain consistent results using 
two different methods to calculate the evidence reliably: the error on
the log evidence differences is conservatively estimated to be of the
order of unity. The details of the evidence calculation method  is
presented elsewhere \citep{beltran}.

We have chosen three datasets for comparison:

\begin{itemize}
\item \CMB: We use the ``standard'' selection of CMB experiments: the
  WMAP data \citep{2003ApJS..148..135Hshort} together with latest VSA
  \citep{2004MNRAS.353..732Dshort}, CBI
  \citep{2004ApJ...609..498Rshort} and ACBAR data \cite{2004ApJ...600...32K}. We also used a
  modified version of the likelihood code that correctly accounts for
  the largest WMAP scales \citep{2004PhRvD..69l3003S}

\item \SN: We use the Riess et al. (2004) SN data. We use both ``gold'' and ``silver''
  datasets. We implemented our likelihood code and checked that it
  gives results consistent with Riess et al. 

\item \SDSS: Finally we use large scale power spectrum measurements
  from the SDSS experiment
  \citep{2000AJ....120.1579Yshort,2002AJ....123..485Sshort,2003AJ....126.2081Ashort}. We
  used the likelihood code by Tegmark \citep{2004PhRvD..69j3501Tshort}
  adapted for \texttt{CosmoMC} by Samuel Leach (private communication).
\end{itemize}

We investigate a 7-parameter cosmological model.  In Table \ref{prior}
we show the uniform priors assumed for the parameters of our model.  We
take our priors to be comparatively broad to approximate the state of
ignorance we may have been in before any of the three experiments were
performed. This has the effect of giving the data as much ``freedom'' as
possible, and correspondingly making the evidence test somewhat
conservative.

\begin{table}
\begin{center}
\caption{The priors assumed for the cosmological model considered in
  this paper. The notation $(a,b)$ for parameter $x$ denotes a top-hat prior in the
range $a<x<b$}
\label{prior}
\begin{tabular}{lc}
\hline\hline 
Basic parameter & Prior \\
\hline
$\omega_{\rm b}$ & $(0.005,0.05)$ \\
$\omega_{\rm dm}$ & $(0.01,0.4)$ \\
$\Omega_{\rm k}$ & $(-0.3,0.3)$ \\
$h$ &  $(0.4,0.9)$ \\
$n_s$ &  $(0.8,1.2)$ \\
$\tau$ & $(0.01,0.7)$\\
$\log 10^{10} A_s$ & $(1,5)$ \\
\hline\hline
\end{tabular}
\end{center}
\end{table}

%-----------------------------------------------------------------------

\section{Results and Discussion}
\label{sect:results}

In Table \ref{tab:results} we give the values of $\Iratio$ for various
combinations of datasets under discussion. We do not detect any
discrepancy between datasets: all combinations of the datasets weakly
favour $\modelA$. 
In the last line of the table \ref{tab:results} we report on the value
of $\Iratio$ for all experiments combined. In principle, it is possible
to have three experiments be pair-wise consistent with each other, but
not when all combined together (imagine, for example three degeneracy
lines forming a triangle). Comfortingly enough, the three-way evidence
test also abrogates $\modelB$ and due to a large number of extra
parameters (\emph{i.e.} twice as many as in other datasets) it has also
a more positive detection of concordance.

\begin{table}
\centering

\caption{The logarithm of $\Iratio$ for various
combinations of datasets. See text for further discussion. }

\begin{tabular}{cc}
\hline

Dataset combination     &       $\log \Iratio$ \\
\hline\hline
CMB -- SDSS             & $0.23 $ \\
SDSS -- SN              & $1.5$ \\
SN -- CMB               & $1.6$ \\ \hline
CMB -- SDSS -- SN & $4.5$ \\ \hline
\end{tabular}

\label{tab:results} 
\end{table}

We have illustrated our methodology with application to real cosmological 
data.  As expected, the data are concordant: any obvious conflict in
the data would likely have been noticed using the ``chi by eye''
methods employed to date.  However, should such discrepancies occur in
the future it is imperative to have a method to quantify these
discrepancies in the most general settings where Gaussianity cannot be
assumed and ever more complex parameter spaces are to be dealt with.

A value of $\Iratio$ less than unity ($\log \Iratio < 0$) is a sign that we should
investigate the mismatch between datasets further. This can be done by
exploring more focussed models, either with new cosmological parameters
(if the experiments are reckoned to be well-understood), or with
additional nuisance parameters that quantify the possible systematic
errors in the data. 
Disentangling the degeneracy between new physics and systematic error
can only be done if the additional parameters come with fresh
information encoded in their prior pdf: this information is then folded
into the evidence ratio, providing the crucial difference between this
methodology and any method relying on goodness-of-fit alone.

%-----------------------------------------------------------------------

\section*{Acknowledgments}

We thank Mike Hobson, Sarah Bridle, Jo Dunkley, Andrew Liddle,
Uro\v{s} Seljak, Antony Lewis for useful discussions.
We also acknowledge Andrew Jaffe for advice on a previous version of
this work.
NR is supported by a PPARC studentship. AS is supported by Ministry of
Education, Science and Sport of the Republic of Slovenia. This work  was
supported in part by the U.S. Department of Energy under contract 
number DE-AC02-76SF00515.

%-----------------------------------------------------------------------

\label{lastpage}
% \bibliography{refs}

\end{document}